\documentclass[11pt]{article}
\usepackage{iap2000,epsfig}

\def\ra{\rightarrow}

\def\be{\begin{equation}}
\def\ee{\end{equation}}
\def\bea{\begin{eqnarray}}
\def\eea{\end{eqnarray}}
\begin{document}
\vspace*{4cm}
\title{Hierarchical Clustering and Active Galaxies.}

\author{ E. Hatziminaoglou (1), G. Mathez (1), A. Manrique (2)}

\address{ (1) Observatoire Midi-Pyr\'en\'ees, Laboratoire 
d'Astrophysique, UMR 5572, 14 Avenue E. Belin, F-31400 Toulouse,
France\\
(2) Dept Astronomia i Meteorologia, Fac de Fisica, Univ. de Barcelona\\
Marti Franques 1, 08028 Barcelona, SPAIN}

\maketitle\abstracts{
The growth of Super Massive Black Holes and the parallel development of
activity in galactic nuclei are implemented in an analytic code of
hierarchical clustering. The evolution of the luminosity function of
quasars and AGN will be computed with special attention paid to the
connection between quasars and Seyfert galaxies. One of the major
interests of the model is the parallel study of quasar formation and
evolution and the History of Star Formation.}

\section{The Barcelona model}

The model of Barcelona has been presented by E. Salvador-Sol\'e in his 
oral contribution entitled ``{\it Modeling the density-morphology 
relation and the galaxy/AGN connection}". A detailed description is also 
given in Manrique et al., 2000, in preparation. Here, only a brief 
introduction will be made, to outline the major differences between the
abovementioned model and previous works.

The big novelty of the model is the use of a {\it modified} Press - Schechter 
formalism of the hierarchical dark matter halo (DMH) formation. 
The internal structure of the dark matter halos is modeled and 
a distinction is made between dark matter halo merging and accretion. 
Next to the dark matter evolution the baryonic matter evolution is also 
followed, through the mechanisms of radiative cooling, star formation
and re-heating, where all component (hot and cold gas, and stars) are
taken into account.

The evolution of galaxies of different types is followed, as well as
their localization within groups and/or clusters.
The evolution of the central galaxy depends on the properties of the
host halo and of the surrounding satellite galaxies that it captures.
This captures depend on the orbits of the galaxies, in other words on the
potential well and the initial orbital conditions. The capture of such
satellite galaxies, which can be very numerous but whose sizes are small 
in comparison to the central galaxy, only produces minor effects. On the
contrary, in the case of a capture of a galaxy with a mass comparable
to the central galaxy, the latter's disk (if it exists) can be destroyed,
forming thus a new spheroid. This kind of capture is, therefore, 
crucial for the final configuration of the central galaxy and for the 
galactic halo gas, due to feedback mechanisms.

Generally speaking, a central galaxy is characterized by its total baryonic
mass, the mass of its gaseous and stellar components and the respective 
metalicities, the star formation rate in the disk and the bulge, the
surface density of the disk and the mass of the central black hole. The 
destiny of the black hole and the AGN activity are related to the history 
of the galaxy. This si what triggers the parallel study of ``normal"
galaxies  and AGN.  

\section{Central Black Hole Evolution and Nourishing Mechanisms}\label{BHevo}

After their formation black holes evolve nourished mainly
by host galaxies interactions.
Galactic bulges can collect material in three different ways:
by cooling flows, by the direct infall of low momentum material from the
galactic 
halo during a merger event between two galaxies of comparable sizes, and
a mass transfer from the disk to the bulge through non-axisymmetric
perturbations (e.g. spiral arms and bars). 

\subsection{ Characteristic Timescales}\label{times}

A certain number of characteristic timescales are involved in the modeling of
black holes and their connection to normal galaxies. Table \ref{tab1}
summarizes some of them, in increasing order. Not all of them appear
in the present paper but all of them are used in the above described model.
``AD" and ``BLR" denote the accretion disk and broad line region, 
respectively.

\begin{table}[ht]
\caption{Characteristic timescales involved in the modeling of AGN}
\label{tab1}
\begin{tabular}{l|cl}
\hline
\hline
& duration & description\\
\hline
\hline
$t_{heat,AD}$ & ? & heating time of the material falling onto the AD\\
$t_{cross,AD}$ & hours - months & time needed by photons to cross the AD\\
$t_{heat,BLR}$ & $\sim$ 1 day & Compton heating time in the BLR\\
$t_{cross,BLR}$ & $\sim$ days & crossing time of the inner BLR by photons\\
$t_{infall,BLR}$ & $\sim$ a few years & gas infall onto the AD through BLR\\
$t_{infall,BH}$ & 10$^3$ -- 10$^5$ years & gas infall from AD to BH\\ 
$t_{Edd}$ & $\sim$ Myr & timescale for the growth of the BH through accretion\\
$t_{dyn}$ & 1 -- 100 Myr & dynamical time of the host galaxy bulge\\
$t_{acc}$ & 3$t_{dyn}$ & quasar ``duty cycle"\\
$t_{cool,BLR}$ & $\sim$ 0.1 -- 10 Gyr & Compton cooling time in the BLR\\
$t_{quiet}$ & a few Gyr & quasar quiescence phase\\
\hline
\hline
\end{tabular}
\end{table}

\subsection{Galaxy Mergers in the DMH Centers}\label{merge}

When galaxies of comparable sizes merge, usually near the center of the 
dark matter halos,
the disks that possibly exist are destroyed and a new spheroid is
formed.  Due to dynamical friction the two black holes (with masses 
$M_{BH}^1$ and $M_{BH}^2$) of the involved galaxies will soon find their way 
towards the central region and coalesce. 
A fraction $\epsilon$ of the cool gas, $ M_{gaz}$, of the spheroid
will fall onto the galaxy center, nourishing the black hole, whose mass
will now be: 
$$M_{BH} = M_{BH}^1 + M_{BH}^2 + \epsilon(f_*M_* + M_{gaz}),$$
where $f_*$ denotes the fraction of the stelar mass accreted, $M_*$, generally
considered to be null.

The typical accretion and radiation time for a black hole is 
nowadays believed to be much shorter than the characteristic timescales
of its host galaxy. However, it is this ``short-term" evolution of the
black hole that determines the light curve of an active galactic nucleus,
and this why it should be modeled. In the case of a rapid growth of the 
bulge we suppose an exponential variation of the accretion rate, inspired
by Dopita (1997), but asymmetrically bell-shaped:
$$\frac{dM_{BH}}{dt} \propto \frac{\Delta M_{BH}}{t_{acc}}\left[1-exp\left
(-\frac{t}{t_{dyn}}\right)\right]exp\left(-\frac{t}{t_{acc}}\right),$$
where $\Delta M_{BH}=\epsilon(f_*M_* + M_{gaz})$ and $t_{acc}=3t_{dyn}$, as
explained in table \ref{tab1}. 

\begin{figure}
\centerline{
\psfig{figure=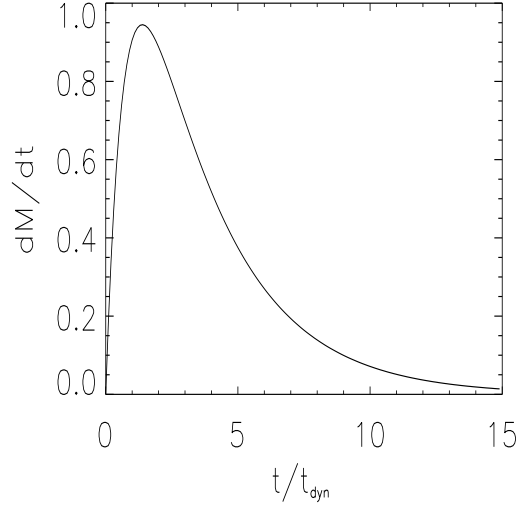,width=8cm,height=8cm}}
\caption{Accretion rate versus time (in units of $t_{dyn}$)}
\label{mdot}
\end{figure}

The difference between this and the Dopita curves is the use of two 
different timescales: $t_{dyn}$ for the ascending part and  $t_{acc}$
for the descending part. This accretion rate, in units of $t_{dyn}$,
is presented schematically in figure \ref{mdot}. The same equation
applies in the case of a black hole nourished through the process
of cooling flows.

\subsection{Slow Fueling}\label{sfuel}

In the case of a slow but continual growth of the galactic bulge as 
spiral arms and bars transfer angular moment towards the outer 
regions and mass towards the center of the galaxy, the mass of the (new)
black hole will be given by: 
$$M_{BH} = M_{BH}^0 + \tilde{\epsilon}(f_*M_*^ {D \ra B} + 
M_{gaz}^{D \ra B}),$$ 
where $M_*^ {D \ra B}$ and $M_{gaz}^{D \ra B}$ denote the stellar and
gaseous mass, respectively, transfered from the disk to the bulge.
In this case a constant accretion rate is adopted:
$$\frac{dM_{BH}}{dt}=\frac{\Delta M_{BH}}{t_{acc}},$$ 
where $\Delta M_{BH}$ is equal, this time, 
to $\tilde{\epsilon}(f_*M_*^ {D \ra B} + M_{gaz}^{D \ra B})$.

\section{AGN Light Curves}

According to the present model, the bolometric luminosity of an AGN
of a given type is determined by its accretion rate, $dM_{BH}/dt$,
the initial mass of the central black hole, $M_{BH}$, 
and the time elapsed since the beginning of the accretion process.
The luminosity sustains the inflow rate and vice versa,
in such a way as to respect the Eddington regime, since
only an under-Eddington luminosity gives a stationary solution, as 
described in Manrique et al. For a given accretion rate
the light emitted from a region of radius $R_{acc}$, and which in fact
is the accretion disk, varies with time as:
$$L_{BH}(t)=\epsilon_{Edd}L_{Edd}={L_{Edd} \over 
{1 + {L_{Edd}R_{acc} \over GM_{BH}}}} \times
\frac{dM_{BH}}{dt},$$
where $L_{Edd}$ is the Eddington luminosity. 

In the case of major mergers or cooling flows, this relation will give 
rise to a bell-shaped light curve with the same 
characteristic timescales as for the accretion rate, $dM_{BH}/dt$.

\section{Conclusions and Perspectives}\label{con}

The model presented above will allow us to make predictions on several
issues that relate quasars to normal galaxies and to the background 
cosmology. 
The problem of the existence of a BH within all 
galaxies and the time delay between
the collapse of a dark matter halo and the formation of a black hole in
its center will be studied first.
The relations between the different types of AGN, their connection to 
normal galaxies, and the relations between
the masses of the dark matter halos, the galactic disks and
bulges, and the black holes can then be looked for.
Issues like the role of AGN and obscured AGN in the reionization history 
of the Universe, the X-ray and UV backgrounds can be examined.

This model will give as an output a theoretical quasar luminosity
function and its evolution with time. Its comparison with the
observed luminosity function
issued from recent or new large quasar samples (e.g. VIRMOS, 2dF, SDSS)
will allow us to better adjust the values of our parameters and tests
our assumptions.

\end{document}